\documentclass[
    ,final            
  ]
  {aipproc}

\layoutstyle{6x9}

\begin{document}

\title{Electroweak physics at HERA\footnote{Talk presented at 9$^{th}$ Conference on the
Intersections of Nuclear and Particle Physics (CIPANP) 2006, Rio Grande, Puerto Rico.}}


\author{Juraj Bracinik (for H1 and ZEUS collaborations)}{
  address={MPI for Physics, Munich}
}

\begin{abstract}
The HERA experiments have collected interesting luminosity samples with both polarized and unpolarized lepton beams. The
polarized CC data are directly sensitive to right handed charged currents. The data are in agreement with the SM
predicting their absence. Effects of Z-exchange have been seen in the lepton beam charge and polarization assymetries.
A combined electroweak and QCD analysis allows one to measure electroweak parameters with world competitive accuracy.
\end{abstract}

\maketitle


\section{Introduction}

HERA is the first and up to now the only electron-proton collider. 
It operates at a total
center-of-mass energy $\sqrt{s}=318 \; {\rm GeV}$, enabling besides 
$\gamma$-exchange a large fraction  of Z and W exchange. Thus HERA is a truly
electroweak collider allowing to study weak interaction effects in the spacelike
region.
The data are collected by the two collider 
experiments, H1 and ZEUS.

Inclusive electron-proton scattering can be conveniently characterized by
three invariant variables, the photon virtuality Q$^2$, Bjorken-x, and inelasticity y. 
At fixed center-of-mass energy only two
of these are independent, as Q$^2$=sxy.

There are two distinct types of interactions  observed at HERA. In the case of
Neutral Current (NC) scattering, either a $\gamma$ or Z-boson is exchanged.
Experimentally such processes are characterized by a scattered electron and a
hadronic jet (or jets) observed in a detector,  balanced in 
transverse momentum. The NC cross section  can be written in
the following form:
\begin{equation}
\frac{d \sigma_{NC}^{e^{\pm}p}}{dx dQ^2} =
\frac{2 \pi \alpha^2}{x Q^4} 
\left[ Y_+ \tilde{F_2} \mp Y_- x \tilde{F_3} - y^2 \tilde{F_L} 
\right] ,
Y_{\pm} = 1 \pm \left( 1-y \right) ^2
\end{equation}
Here the unknown proton
structure is expressed  in terms of three generalized structure functions. The 
dominant contribution to the cross section is provided by $\tilde{F_2}$; in the
quark-parton model (QPM) it is equal to the  sum of parton density
functions
(pdfs) of quarks and antiquarks weighted by their charges squared.  

At large momentum
transfers $x\tilde{F_3}$ becomes sizable, in QPM it is proportional to the  sum of pdfs of valence
quarks weighted by their charges squared. The longitudinal structure function  $\tilde{F_L}$ is visible only at
high y and can be neglected in this article.

In the case of Charged Current (CC) scattering a W boson is exchanged and
the electron is transformed to a neutrino. Such events are experimentally
characterized by a hadronic jet (or jets) with missing transverse momentum.
Expressing the structure of the proton in terms of the functions $W_2$, $xW_3$ and $W_L$,
the CC cross section can be
parametrized as:
\begin{equation}
\frac{d \sigma_{CC}^{e^{\pm}p}}{dx dQ^2} =
\frac{G_F^2}{4 \pi x} \left[
\frac{M_{{\rm prop}}^2}{M_{\rm prop}^2 + Q^2}
\right]^2
\left[
Y_+ W_2 \mp Y_- x W_3 - y^2 W_L 
\right]
\label{eq:cc1}
\end{equation}
  Here M$_{prop}$ is the mass of the exchanged boson occuring in the propagator.  
\section{EW effects in unpolarized NC/CC scattering}  
\begin{figure}
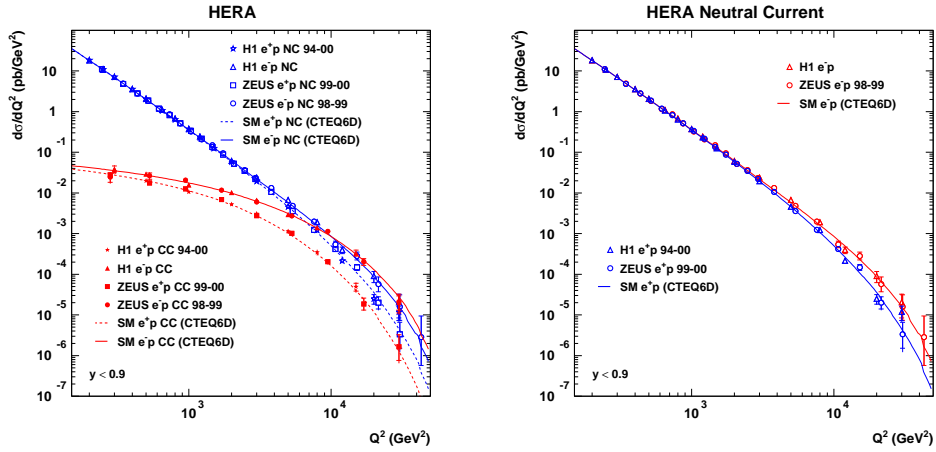

  \includegraphics[height=.3\textheight]{nc_cc_dsdq2_cteq6d.eps.gv}
   \includegraphics[height=.3\textheight]{nc_dsdq2_cteq6d.eps.gv}
  \caption{Differential cross sections of unpolarized NC and CC scattering as a
  function of Q$^2$ for NC and CC (left) and for $e^{\pm}p$ NC (right).}
  \label{fig:fig1}
\end{figure}
The NC and CC differential cross sections as a function of Q$^2$ are
shown on Fig.~\ref{fig:fig1} left. While they differ by several orders of magnitude at low 
Q$^2$, due to different propagator terms, at Q$^2$ close to the W-mass squared
they are of similar magnitude. The remaining differences are caused mainly by
the fact, that NC and CC are sensitive to different combinations of pdf's.

On Fig.~\ref{fig:fig1} right one can see the differential  NC cross section as a function of
Q$^2$ for $e^+p$ and $e^-p$ collisions. At high Q$^2$ the dependence on 
the lepton beam charge is visible. This effect is described by
 $x\tilde{F_3}$, its main contribution at
HERA energies being the $\gamma$Z interference.
  
\section{Polarized NC/CC scattering}
\begin{figure}
  \includegraphics[height=.3\textheight]{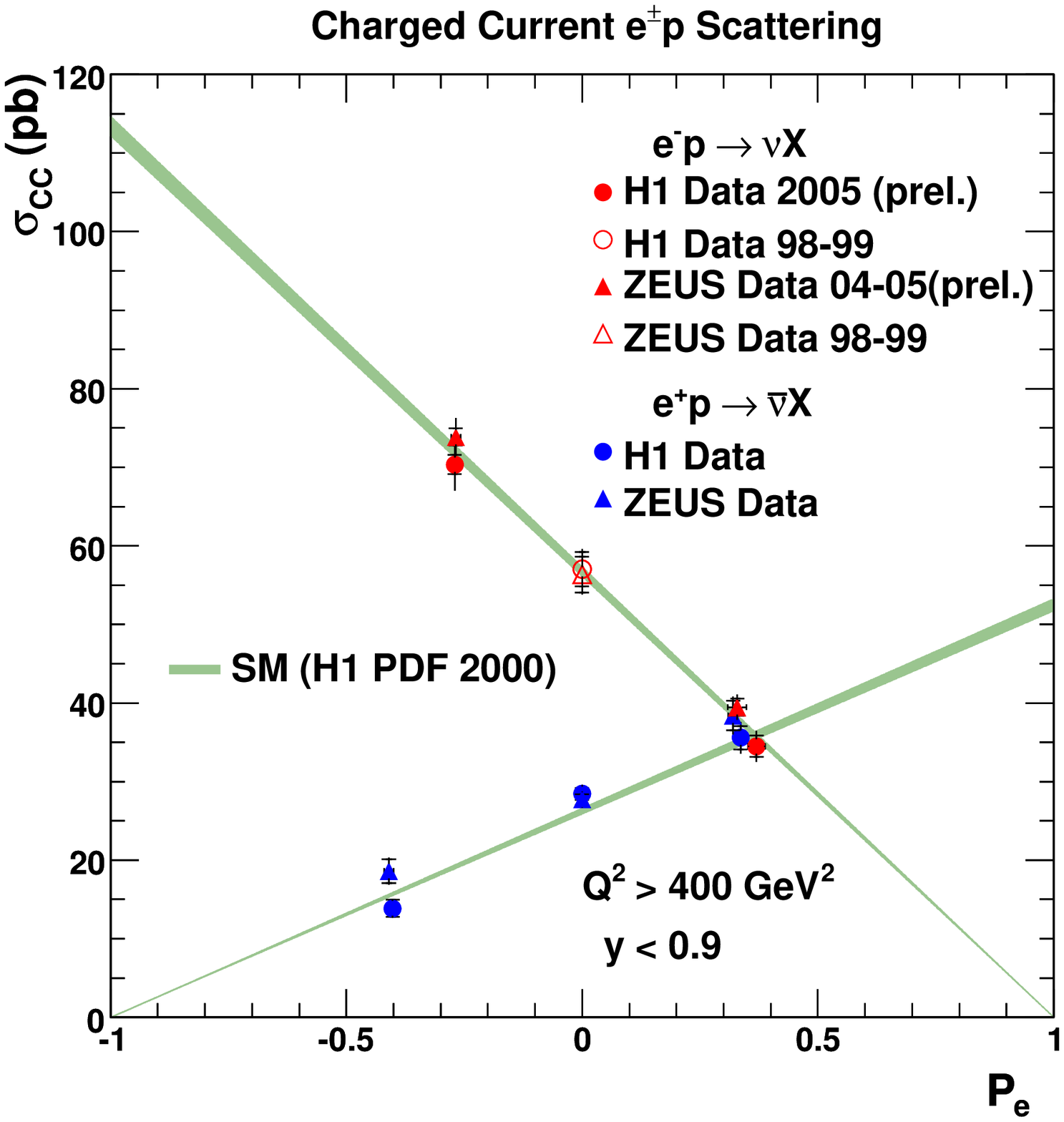}
  \includegraphics[height=.29\textheight]{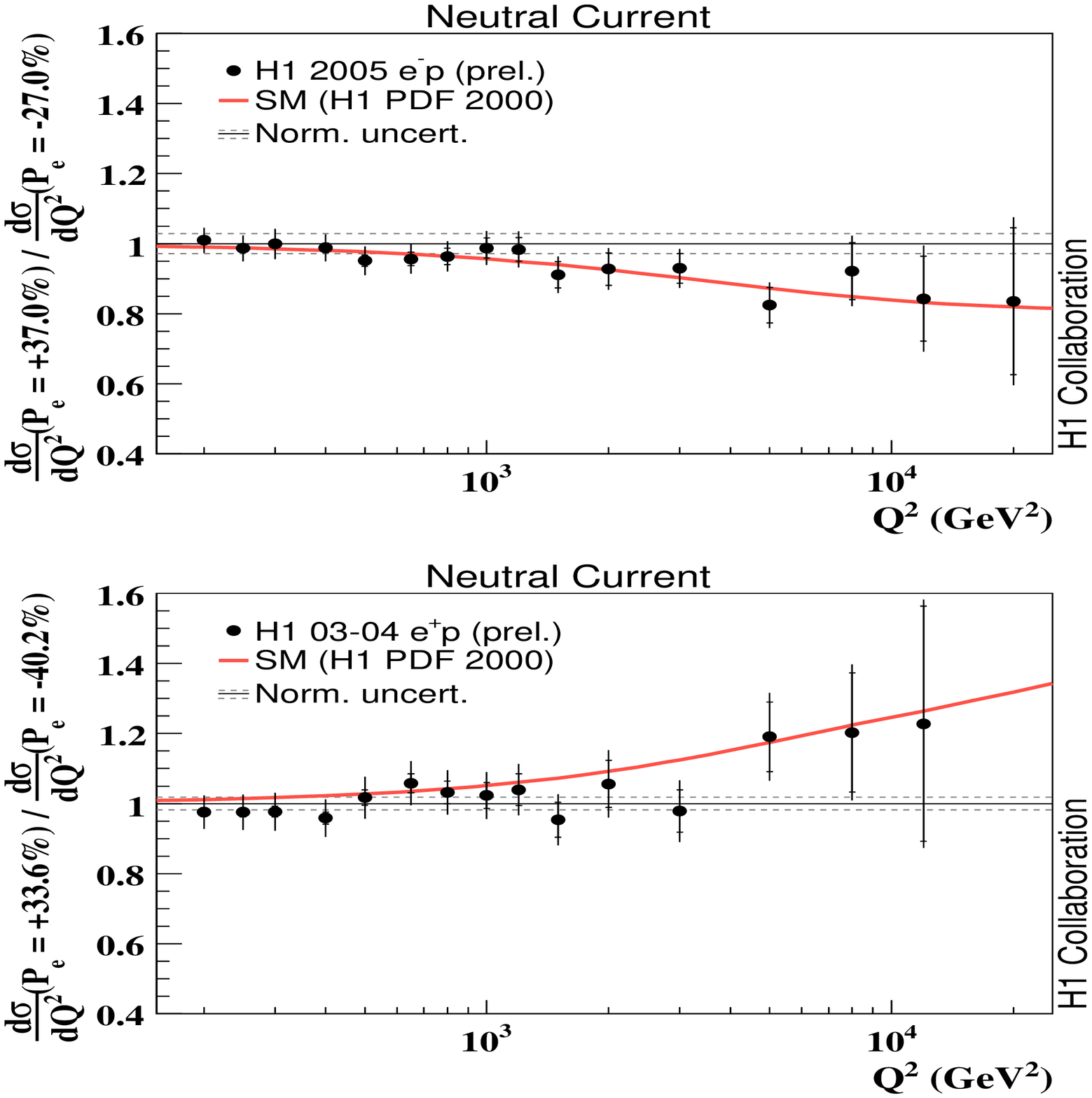}
  \vspace{-1.5cm}
  \caption{Differential cross sections of polarized  CC as a function of
  pollarization (left) ad ratio between NC cross sections for two polarizations for
  electron an positron beam (right).}
  \label{fig:fig2}
\end{figure}
Starting in 2000, the HERA collider was upgraded to deliver  higher
luminosities and longitudinally polarized lepton beams.

The transverse polarization in the HERA electron ring builds up naturally via
emision of synchrotron radiation (Sokolov-Ternov effect). Spin
rotators flip the transverse polarization to longitudinal just before the interaction
regions and vice versa behind it. The level of longitudinal polarization achieved up
to now is 30-40 \% .

The longitudinal polarization has a particularly strong effect on charged current
cross sections. In the Standard Model (SM) only left handed particles (right handed
antiparticles)  interact via CC. Thus, defining the level
of polarization as $P_e=(N_R-N_L)/(N_R+N_L)$, where $N_L$ and $N_R$ are  
numbers of left handed (right handed) leptons in the beam, a linear dependence of the
cross section on $P_e$, giving zero cross section at $P_e=1(-1)$ for
electron (positron) beam is expected.

Indeed, the total CC cross section as a function of
$P_e$, integrated over the visible phase
space indicated on Fig.~\ref{fig:fig2} left, is in good agreement with the SM prediction. 

A contribution of a right-handed CC should change the slope and the intercept of the 
linear dependence of $\sigma_{CC}$ on $P_e$. The cross sections were fitted by  straight lines and
extrapolated to $P_e=1(-1)$ for
electron (positron) beam. The H1 collaboration has presented data, for the visible
phase space Q$^2$ > 400 GeV$^2$ and y< 0.9, leading to $\sigma_{CC}(e^+p, P_e=-1) = -3.9 \; 
 \pm 2.3 ({\rm stat}) \pm 0.7 ({\rm sys}) \pm 0.8 ({\rm pol}){\rm pb}$ and 
$\sigma_{CC}(e^-p, P_e=+1) = -0.9 \; 
 \pm 2.9 ({\rm stat}) \pm 1.9 ({\rm sys}) \pm 2.9 ({\rm pol}){\rm pb}$. The ZEUS collaboration 
has shown results, for the  visible phase space Q$^2$ > 200 GeV$^2$ leading to $\sigma_{CC}(e^+p, P_e=-1) =
7.4 \;  \pm 3.9 ({\rm stat}) \pm 1.2 ({\rm sys}){\rm pb}$ and
$\sigma_{CC}(e^-p, P_e=+1) =
0.8 \;  \pm 3.1 ({\rm stat}) \pm 5 ({\rm sys}){\rm pb}$. All results are 
compatible with zero, as predicted by the SM.
%
%

For NC scattering, both $\tilde{F_2}$ and $x\tilde{F_3}$ 
depend on the
polarization due to pure Z exchange and $\gamma$-Z interference: 
\begin{equation}
\tilde{F_2} = F_2 - \left( {\rm v}_e \pm P_e {\rm a}_e 
\right) \chi_{\rm Z} F_2^{\gamma {\rm Z}} + \left(
{\rm v}_e^2 + {\rm a}_e^2 \pm 2 P_e {\rm v}_e {\rm a}_e
\right) \chi_{\rm Z}^2 F_2^{\rm Z}    
\end{equation}
\begin{equation}
x\tilde{F_3} = - \left({\rm a}_e \pm P_e {\rm v}_e
\right) \chi_{\rm Z} x F_3^{\gamma {\rm Z}} +
\left(2 {\rm v}_e{\rm a}_e \pm P_e({\rm v}_e^2+{\rm a}_e^2)
\right) \chi_{\rm Z}^2 x F_3^{\rm Z} 
\end{equation}
where $\chi_{\rm Z}$ is the ratio of two propagators, 
$ \chi_{\rm Z} = 1/(\sin ^2 2 \theta_{\rm W}) Q^2/(M_{\rm Z}^2+Q^2)$, and ${\rm v}_e$ 
and ${\rm a}_e$
are vector and axial couplings of Z to electron. At HERA energies, the contributions 
of $F_2^{\rm Z}$ and $x F_3^{\rm Z} $ are
suppressed by the propagator. As ${\rm v}_e << {\rm a}_e$, 
to first order $\tilde{F_2} \approx F_2 \pm P_e {\rm a}_e \chi_{\rm Z} F_2^{\gamma {\rm Z}}$ 
and therefore one expects the polarization asymmetry to be of
similar size but opposite sign for $e^+p$ and $e^-p$ interactions. 
This is indeed observed in the  ratio of NC cross sections, taken with two slightly
different polarizations, for  $e^+p$ and $e^-p$ as shown on Fig.~\ref{fig:fig2} right. 
  
\section{Combined EW and QCD analysis of  NC/CC data }
\begin{figure}
 \includegraphics[height=.3\textheight,angle=-90.]{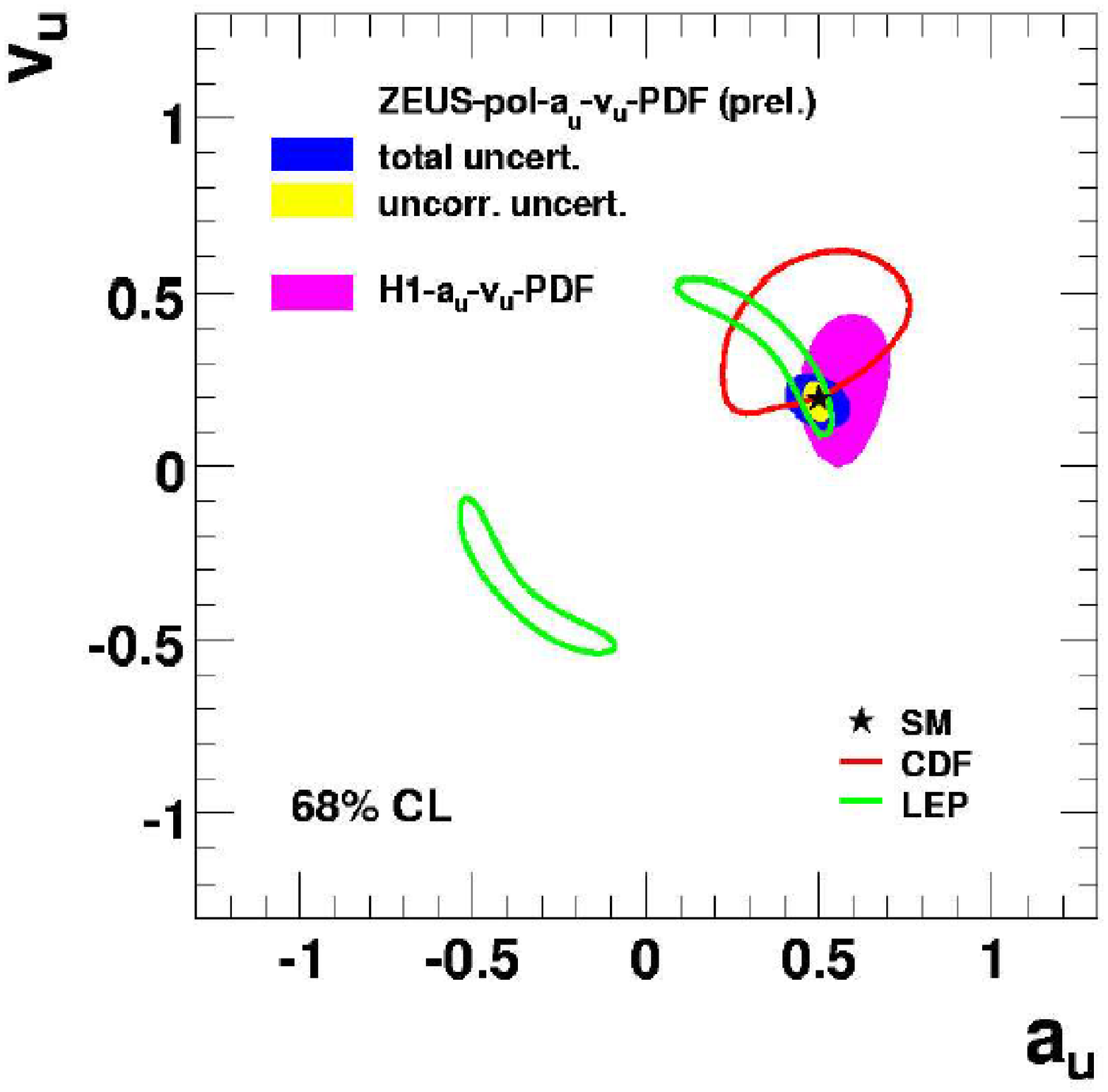}
\includegraphics[height=.3\textheight,angle=-90.]{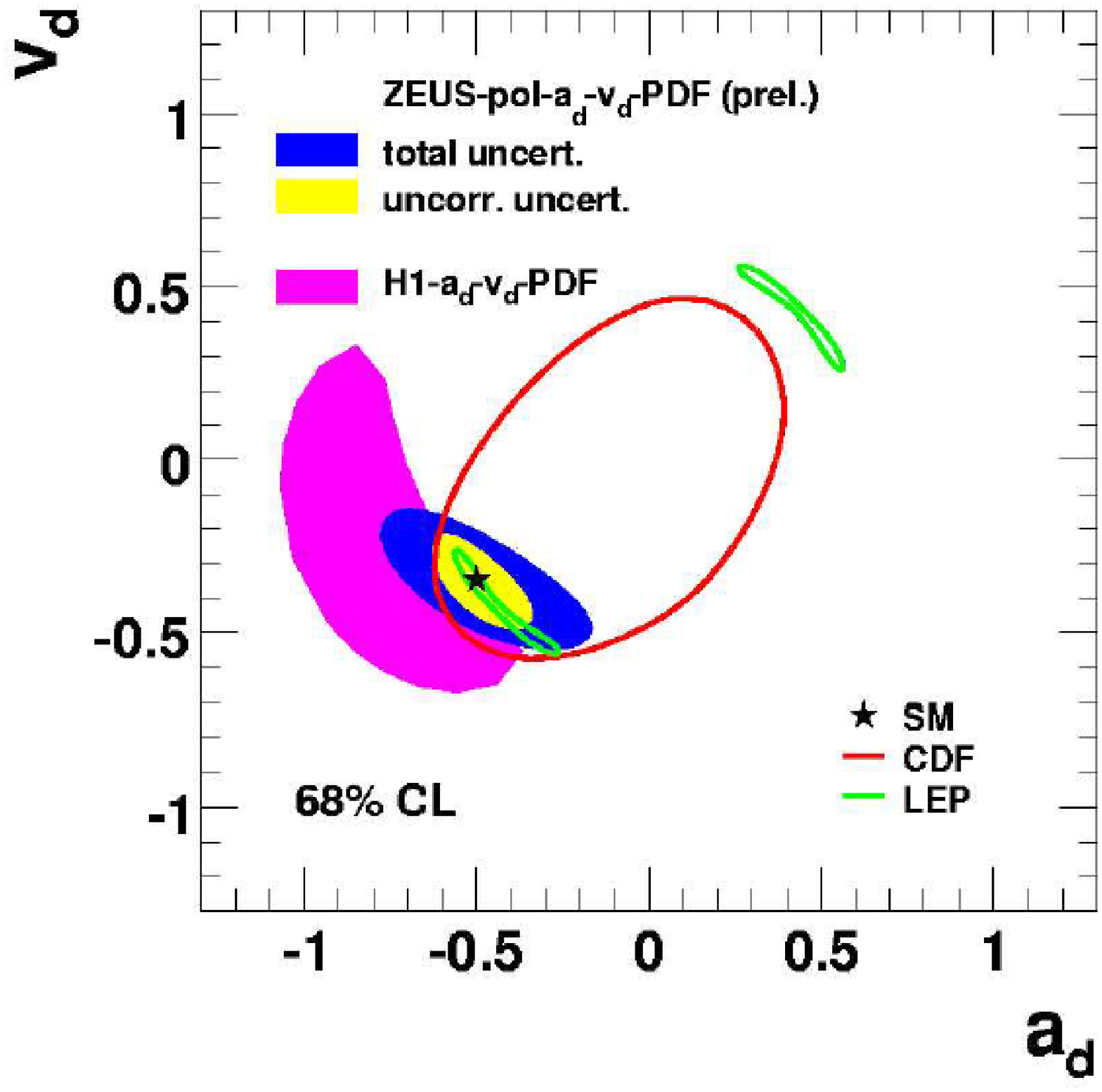}
  \caption{Measured axial and vector couplings of light quarks to Z-boson}
  \label{fig:fig4}
\end{figure}
Good statistical precision and wide phase space coverage of HERA data allow one to perform a
combined analysis, where the QCD 
and electroweak parameters are determined
simultaneously from a fit to the data. The H1
analysis made use of HERA I unpolarized data, the preliminary ZEUS analysis included HERA II data.


The NC data are sensitive to the vector and axial coupling of light quarks to the Z-boson. 
In the QPM both 
$\tilde{F_2}$ and $x\tilde{F_3}$ can be expressed as: 
$F_2^{\gamma {\rm Z}} = 2 \sum e_q {\rm v}_q x[q + \overline{q}]$, 
$F_2^{\rm Z} =  \sum ({\rm v}_q^2 + {\rm a}_q^2) x[q + \overline{q}]$, 
$xF_3^{\gamma {\rm Z}} = 2 \sum e_q {\rm a}_q x[q - \overline{q}]$ and 
$xF_3^{\rm Z} = 2 \sum {\rm v}_q {\rm a}_q x[q - \overline{q}]$, 
where $e_q$ is the quark electric charge. 

Using ep data, it is possible to determine both axial and vector couplings simultaneously, 
as contributions from Z-exchange and the
 $\gamma$Z-interference have different Q$^2$ dependence. Thanks to the interference terms, 
 there is no sign ambiguity. Inclusion of the
polarized data improves the determination of 
${\rm v}_q$.

The results are shown in Fig.~\ref{fig:fig4}. They are in good agreement with the 
Standard Model and competitive with 
measurements from LEP and CDF.


\begin{theacknowledgments}
  I would like to thank G. Grindhammer, M. Klein, K. Nagano and
  E. Perez for usefull discussions.
\end{theacknowledgments}



\bibliographystyle{aipproc}   




\end{document}